\documentclass[amsmath,amssymb]{revtex4}

\usepackage{graphicx}
\usepackage{dcolumn}
\usepackage{bm}

\begin{document}

\title{Interfacial structure and electronic properties of AFM/FM magnetic multilayer revealed by Scanning Tunneling Microscope}

\author{S. Nakashima}
\affiliation{The Institute for Solid State Physics, The University of Tokyo, Kashiwa, Chiba 277-8581, Japan}
\author{T. Miyamachi}
\affiliation{The Institute for Solid State Physics, The University of Tokyo, Kashiwa, Chiba 277-8581, Japan}
\author{F. Komori}
\affiliation{The Institute for Solid State Physics, The University of Tokyo, Kashiwa, Chiba 277-8581, Japan}
\date{\today}

\begin{abstract}
Mixing of atoms at the interface was studied for Mn/Fe magnetic hetero-epitaxial layers on Cu(001) by scanning tunneling microscopy/spectroscopy. The formation of a surface alloy was observed when the Mn layer was thinner than 3 atomic layers. From the fourth layer, Fe segregation is suppressed, and a pure Mn surface appears. Accordingly, spectroscopic measurements revealed the electronic difference between the surface alloy and Mn layers. The surface electronic structure of the fourth Mn layer is slightly different from that of the fifth layers, which is attributed to the hybridization of the fourth layer with the underneath Fe-Mn alloy.
\end{abstract}
\maketitle

\section{Introduction}
Utilizing the magnetic coupling between two magnetic layers has been the focus of interest for magnetic multilayer devices \cite{1988PRL_Baibich,2001Science_Wolf,2008springer_Zabel}. The fundamental magnetic properties of magnetic multilayers, e.g., magnetic moment, magnetic anisotropy and coercivity, rely much on their interfacial structural and electronic properties. Importance of the interface quality is especially highlighted for the multilayer spin-valve system\cite{1994MMM_Dieny,1999MMM_Nogues,2005PhysRep_Nogues}. In a ferromagnetic (FM) multilayer system, through a non-magnetic spacer layer in between the two FM layers, the magnetically free FM layer interacts with the other FM layer whose magnetization is pinned by the exchange coupling with an antiferromagnetic (AFM) layer. The magnitude of the exchange bias or the spin-dependent transport properties could be significantly degraded by the local disordered structures such as atomic defects, roughness, intermixing, which modify electronic structures at interface between the magnetic layers \cite{2006NatureMat_Kuch,2003JPCM_Tsymbal,2004PRL_Tiusan}. 

Scanning tunneling microscopy/spectroscopy (STM/STS) is a powerful tool to investigate structural and electronic properties of multilayer films on an atomic scale. In the vicinity of the AFM/FM interface, previous STM works could relate the surface morphology and electronic structures of AFM thin films grown on a FM single crystal to their magnetism 
\cite{2002SS_Bischoff,2002SS_Yamada,2003PRL_Yamada,2004PRL_Schlickum,2006PRB_Schlickum}. 
However, the atomic-scale investigation of the growth, electronic and magnetic properties of the AFM/FM interface in the form of multilayers has rarely done so far.

In this letter, we investigate the influence of interfacial intermixing on the growth and electronic properties of the AFM/FM double-layer system by STM/STS. An fcc Fe thin film on the Cu(001) substrate\cite{1994SS_Schmailzl,2001PRL_Biedermann,2001PRL_Qian} was selected as the FM layer, and Mn films grown on top of that as the AFM layer. Note that, in the Fe fcc phase, the top two Fe layers couple ferromagnetically, while the details of magnetic structures below the third layer are still under debate 
\cite{2001PRL_Qian,2009PRL_Meyerheim,2010PRB_Sandratskii,2012PRB_ViolBarbosa}.

\section{Experimental}

All the STM/STS measurements were performed in an ultrahigh vacuum ($<$ 2.0 $\times$10$^{-11}$ Torr) at 80 K using electrochemically-etched W tips. The samples were prepared in a UHV preparation chamber separated from the STM/STS measurement chamber by a gate valve. A clean and flat surface of Cu(001) was prepared by several cycles of Ar$^{+}$ sputtering and subsequent annealing at 720 K. Pure Fe (purity 99.998$\%$) and Mn (purity 99.99$\%$) were deposited at room temperature by molecular beam epitaxy (MBE) with the growth rate of 0.8 monolayer (ML)/min. Here, ML is defined as the atom density of the clean Cu(001) surface. During the deposition, the base pressure of the preparation chamber was better than 1.0 $\times$ 10$^{-10}$ Torr. The thickness of the Fe films was fixed to be 8 ML. To investigate interfacial mixing between Mn and Fe thin films, the surface morphology was characterized by STM and the local electronic structures were investigated by STS. The topographic images were obtained using a constant tunneling current ($I_{\rm T}$) mode at constant sample-bias voltages ($V_{\rm S}$). The differential conductance of tunneling current, d$I$/d$V$, was recorded using a lock-in technique with a modulation voltage of 20 mV and frequency of 733 Hz. 

\section{Results and discussion}

\begin{figure}[tb]
\begin{center}
\includegraphics[width=80mm]{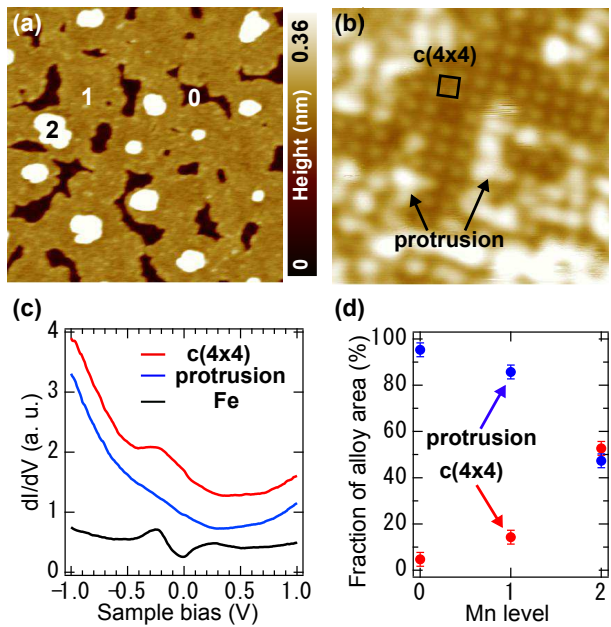}
\end{center}
\caption{(a) STM image of the 8 ML Fe film on Cu(001) with deposited 1 ML Mn (100 $\times$ 100 nm$^{2}$, $V_{\rm S}$ = +1 V, $I_{\rm T}$ = 1 nA). The numbers indicate the local Mn thickness. (b) High resolution STM image on the first layer (10 $\times$ 10 nm$^{2}$, $V_{\rm S}$ = +0.05 V, $I_{\rm T}$ = 3 nA). The square and arrows indicate the $c$(4 $\times$ 4) unit cell and protrusions respectively. (c) d$I$/d$V$ spectrum recorded on the alloy area and the Fe surface before Mn deposition ($V_{\rm S}$ = +1 V, $I_{\rm T}$ = 5 nA). (d) Fraction of the  MnFe alloy area in the various Mn layer levels.}
\label{fig1}
\end{figure}

Figure 1(a) shows an STM image after deposition of 1 ML Mn on the 8 ML Fe film. Before the Fe surface (level 0) was completely covered by the first Mn layer (level 1), the second Mn layer (level 2) started to grow. A high resolution STM image for the first Mn layer (level 1) is shown in Fig. 1(b). Well-ordered areas with a $c$(4 $\times$ 4) reconstructed structure (marked by square), and disordered protrusions with an average height of 20 pm were observed on the surface (marked by arrow). Previous STM studies of the Mn thin films grown on Cu(001) have revealed that Mn-based surface alloys can make a variety of reconstructed structures with different periodicities. Depending on growth conditions, the MnCu surface alloys exhibit $c$(2 $\times$ 2), $c$(8 $\times$ 2), $c$(12 $\times$ 8) and $p$2$m$$g$(4 $\times$ 2) reconstructed structures
\cite{1992SS_Flares}. 
Thus, the observed $c$(4 $\times$ 4) surface reconstruction and the protrusion could be the presence of the MnFe surface alloys with different compositions.

The formation of the surface alloys is also confirmed by measuring the local electronic states. Figure 1(c) presents the d$I$/d$V$ spectra recorded on the areas with the $c$(4 $\times$ 4) reconstruction and with the protrusions. While the spectrum on the $c$(4 $\times$ 4) reconstructed structure shows a broad peak structure around $-$0.2 V, the one on the protrusions shows no characteristic feature. Furthermore, both spectra are considerably deviated from the reference spectrum of the 8 ML Fe films on Cu(001). Thus, these spectroscopic measurements support our argument that the $c$(4 $\times$ 4) reconstructed structure and protrusion are MnFe surface alloys but have different Mn and Fe compositions.

To achieve detailed information on the compositions of observed surface alloys, we first investigate the fractions of the areas with the $c$(4 $\times$ 4) reconstruction and with the protrusions at each Mn layer level. Figure 1(d) shows the fractions of the areas in the bottom (0), first (1) and second (2) layer levels. The fraction of the $c$(4 $\times$ 4) MnFe alloy increases with increasing the Mn layer level. The observed tendency can be interpreted in terms of the segregation of Fe atoms during the Mn deposition. Thermodynamically, less Fe atoms would segregate to the surface as more Mn atoms are deposited. The observed continuous increase in the fraction of the $c$(4 $\times$ 4) MnFe alloy with increasing the Mn layer level thus suggests that the density of the Mn atoms in the $c$(4 $\times$ 4) reconstructed structure is larger than in the protrusion area. 

\begin{figure}[tb]
\begin{center}
\includegraphics[width=80mm]{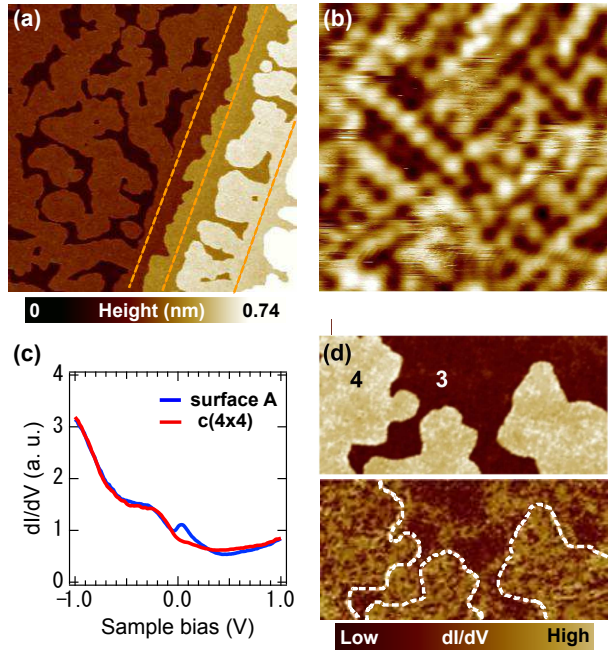}
\end{center}
\caption{(a) Large scale STM image of the 3.7 ML Mn on 8 ML Fe/Cu(001) (240 $\times$ 240 nm$^{2}$, $V_{\rm S}$ = +1 V, $I_{\rm T}$ = 1 nA). Orange dashed lines indicate the position of Fe/Cu step edges. (b) STM image in the $c$(4 $\times$ 4) alloy region on the third layer (10 $\times$ 10 nm$^{2}$, $V_{\rm S}$ = +0.05 V, $I_{\rm T}$ = 5 nA). (c) Two different d$I$/d$V$ curves measured on the third layer: the red curve was measured on the disorder $c$(4 $\times$ 4) area and the blue curve was measured on the surface A ($V_{\rm S}$ = +1 V, $I_{\rm T}$ = 5 nA). (d) (upper panel) STM image and (lower panel) corresponding d$I$/d$V$ map at +0.1 V (120 $\times$ 60 nm$^{2}$, $V_{\rm S}$ = +0.1 V, $I_{\rm T}$ = 3 nA).}
\label{fig2}
\end{figure}

Figure 2(a) presents an STM image after depositing 3.7 ML Mn on the 8 ML Fe film. The Mn film consists of 3 and 4 ML Mn layers. Orange dashed lines represent positions of the step edges of the Fe/Cu(001) substrate. The Mn layers grow in a step-flow and layer-by-layer mode above 3 ML as in the case of the Mn layers on Co films
\cite{2015PRB_Kuch}.
On the third Mn layer, we find two surface structures. One is the disorder $c$(4 $\times$ 4) structure as shown in Fig. 2(b), and an atomic image of the the other surface was not obtained. We call this surface, surface A. The d$I$/d$V$ spectra recorded on third layers are shown in Fig. 2(c). On the disorder $c$(4 $\times$ 4) area, the spectra with a broad peak structure at $-$0.2 V is the same as that on the $c$(4 $\times$ 4) area of the first and second layers [see. Fig. 1(c)]. The spectrum on the surface A has a peak structure at +0.1V with a broad peak structure at $-$0.2 V. On the forth layer, the latter spectra were observed on most of the surface, and there were very few areas with the $c$(4 $\times$ 4) structure. In Fig. 2(d) the d$I$/d$V$ map at +0.1 V demonstrates the distribution of the spectra as in Fig. 2(c). The homogeneous electronic structures of the fourth Mn layer can be seen as a bright contrast in this map, while the bright contrast on the third layer covers about 30 $\%$ of the surface.

\begin{figure}[tb]
\begin{center}
\includegraphics[width=80mm]{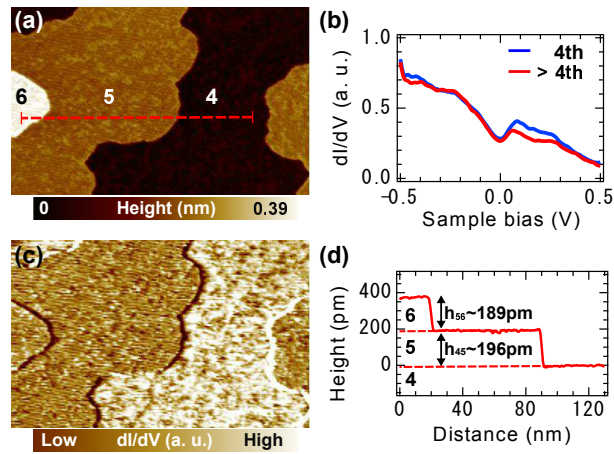}
\end{center}
\caption{STM image of 5 ML Mn on 8 ML Fe/Cu(001) (160 $\times$ 100 nm$^{2}$, $V_{\rm S}$ = +0.1 V, $I_{\rm T}$ = 3 nA). (b) d$I$/d$V$ curves measured on the fourth layer (blue) and above the fourth layer (red) ($V_{\rm S}$ = +0.5 V, $I_{\rm T}$ = 5 nA). (c) d$I$/d$V$ map at +0.1 V at the same area as (a). (d) Line profile along the red dashed line in (a) at +0.5 V.}
\label{fig3}
\end{figure}

Figure 3(a) presents an STM image after deposition of 5 ML Mn on 8 ML Fe/Cu(001). We find three kinds of the Mn layers levels, fourth, fifth and sixth. The d$I$/d$V$ spectrum recorded on this surface shown in Fig. 3(b) reveals that the electronic structures of the fifth Mn layer are quite similar, but the peak intensity located at +0.1 V is slightly weaker compared to those of the fourth Mn layer. Note that the d$I$/d$V$ spectra of Mn layers thicker than 5 ML is almost identical to that of the fifth Mn layer (not shown here). The electronic difference between fourth, fifth, and sixth Mn layers can be clearly seen in the d$I$/d$V$ map at +0.1 V, where the d$I$/d$V$ intensities of the fifth and sixth Mn layers are comparable but weaker than the fourth Mn layer [see Fig. 3(c)]. 

The difference in the electronic structures between the fourth and fifth Mn layers is at a glance strange because of their almost identical surface structure. We attribute this to the difference in the underneath layer of the fourth and fifth Mn layers, i.e., the MnFe surface alloy and pure Mn layer (the fourth Mn layer), respectively. The fourth Mn layer hybridizes with underlying MnFe surface alloy which has no characteristic electronic states at +0.1 V [see. Fig 2(b)]. This could result in the localized electronic structures of the fourth Mn layer around this energy region, and thus stronger peak intensity compared to the fifth Mn layer
\cite{2012NatureCom_Miyamachi}. 
We also notice that the 1 ML height between the fifth and sixth Mn layer is slightly different. Figure 3(d) shows the line profile along the red dashed line in Fig. 3(a). The heights of the fifth and sixth Mn layers, h$_{45}$ and h$_{56}$, are 196 and 189 pm, respectively. Taking the almost identical electronic structures between the fifth and sixth Mn layers into consideration, the observed height difference could be caused by the different layer spacing between h$_{45}$ and h$_{56}$, which is triggered by electronically modified fourth Mn layer.

To get an overall picture of the interfacial condition of Mn/Fe magnetic layers, we finally plot the Mn step height between various layers and present the schematic model of the system as shown in Fig. 4. Note that the number of the Mn layer j is defined as the 1 ML step height between level j-1 and level j. The above STM/STS observations clarify that the first layer is Fe-dominant MnFe surface alloy in the form of the protrusion, which results in its step height close to the that of the Fe fcc(001) lattice of about 180 pm. As the number of the Mn layer increases up to the second and third layers, the composition of the MnFe surface alloy becomes Mn-dominant. Thus, the gradual increase in the step height could reflect the characteristic of Mn atoms. From the fourth layer, the suppressed segregation of Fe atoms allows the growth of a pure Mn layer, and further increases the step height. Above the sixth layer, the step height is a nearly constant of 189 pm. This value is consistent with the interlayer spacing of the face-centered-tetragonal (fct) Mn film on Co/Cu(001) system
\cite{2006APL_Kohlhepp, 2014PRB_Wang}. 
The maximum layer expansion was observed at the surface of the fifth layer.

\begin{figure}[tb]
\begin{center}
\includegraphics[width=100mm]{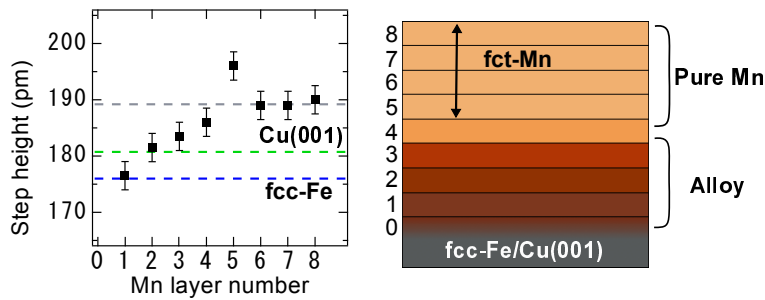}
\end{center}
\caption{(left panel) Step height of the Mn layers as a function of Mn thickness. The step heights between various layers were observed for the sample voltage between $-$1 V and +1 V with 0.1 V interval, and averaged. (right panel) Schematic model of Mn films on Fe/Cu(001).}
\label{fig4}
\end{figure}


\section{Conclusions}

In summary, we have investigated the interfacial structural and electronic properties of the Mn/Fe multilayers on Cu(001) in terms of intermixing with STM/STS. We observe the formation of the surface alloy up to 3 ML Mn. The composition of Fe atoms was found to decrease with increasing the number of Mn layers increases, which results in the presence of two types of the MnFe surface alloys. From the fourth layer, the segregation of Fe atoms is suppressed and the pure Mn layer is formed. However, because of the hybridization with underneath MnFe surface alloy, we find that the electronic structures of the fourth layer are deviated from those of the fifth layer. We conclude from the evaluation of the 1 ML step height at each Mn layer that the growth of the pure Mn layer with the fct structure is promoted from the fifth layer. The present results demonstrate that the overall influence of the interfacial intermixing on the structural and electronic properties of magnetic multilayer systems can be clarified by in-depth STM/STS work.

  
\section{Acknowledgements}
This work is partly supported by JSPS KAKENHI for Young Scientists (A), Grant No. 16H05963, for Scientific Research (B), Grant No. 26287061, the Hoso Bunka Foundation, Shimadzu Science Foundation and Iketani Science and Technology Foundation.

\end{document}